# Relevance Scoring of Triples Using Ordinal Logistic Classification

## The Celosia Triple Scorer at WSDM Cup 2017

Nausheen Fatma
nausheen.fatma@research.iiit.ac.in

Manoj K. Chinnakotla
manojc@microsoft.com

Manish Shrivastava
m.shrivastava@iiit.ac.in

## ABSTRACT

In this paper, we report our participation in the Task 2: Triple Scoring of WSDM Cup challenge 2017. In this task, we were provided with triples of "type-like" relations which were given human-annotated relevance scores ranging from 0 to 7, with 7 being the *"most relevant"* and 0 being the *"least relevant"*. The task focuses on two such relations: *profession* and *nationality*. We built a system which could automatically predict the relevance scores for unseen triples. Our model is primarily a supervised machine learning based one in which we use well-designed features which are used to a make a Logistic Ordinal Regression based classification model. The proposed system achieves an overall accuracy score of 0.73 and Kendall's tau score of 0.36.

## 1. INTRODUCTION

In the recent years, the semantic web and the linked data has gained a lot of popularity for organizing "knowledge" in the form of entities and their relationships. A significant effort has been to put by many communities to develop these large scale knowledge bases like DBpedia [1], YAGO [2], Wikidata [3], etc. Various commercial search engines companies like Google, Microsoft, IBM, etc. are creating their own knowledge bases, and are widely employing it in their applications [1,2]. It is believed to have brought a breakthrough advancement in the search community.

Although it is very simple to query a knowledge base and get a result-set, the result set by itself doesn't have any rank or order. Ranking or scoring the result set is a very challenging and an important area of research. For example, if we query to list the profession for the entity *Barack Obama*, we get a result-set: {*author, lawyer, law professor, politician*}. Applying our general knowledge, we can say intuitively that for *Barack Obama*, the profession *politician* is of a much higher relevance than the profession *author*.

In this paper, we participate in the WSDM Cup challenge [4] to address this task of predicting relevance scores for triples denoting "type-like" relation, for two types of relations : *profession* and *nationality*. We make a system which, for the given triple, predicts an integer score in the range 0 to 7, where 7 denotes *most relevant* and 0 denotes *least relevant*.

The rest of the paper is divided into following sections: Section 2 describe the related work in this area. Sections 3 explains our model and different approaches in detail. Section 4 contains the experimental setup. Section 5 discusses the evaluation results and Section 6 comprises of the conclusion and future work.

---

[1]https://www.google.com/intl/es419/insidesearch/features/search/knowledge.html

[2]https://www.bing.com/partners/knowledgegraph

## 2. RELATED WORK

Ranking and relevance scoring has been a well studied area of research for several years in the Information Retrieval and the World Wide Web community [5, 6, 7]. However, it has not much been explored for data represented in the form of knowledge base triples. Bast *et al.* [8] study the problem of computing relevance scores for Knowledge base triples from type-like relations which essentially measures the degree to which an entity belongs to a given type. The difference from their work is that they have used only word counts/frequency based approaches ranging from rule-based approaches like "First" and "Prefixes", to probabilistic and statistical approaches like "Words MLE" and "Labeled LDA" and learning based approach like "Words Classification". Whereas our approach uses meaning based similarity based features using entity embeddings to build the model. Hogan *et al.* [9] developed and algorithm-*ReConRank* in which they adapted the popular PageRank algorithm for the RDF data. Elbassuoni *et al.* [10] proposed a language model based ranking of the RDF graphs. Fatma *et al.* [11] have studied the problem of mining trivia from knowledge graphs by classifying the triples into two categories: *Boring* and *Interesting*.

## 3. CELOSIA TRIPLE SCORER (CTS)

In this section, we describe the details of our system Celosia Triple Scorer (CTS). Figure 1 shows the architecture of CTS. The white arrows describes the training and the green arrows describes the prediction. During the training, CTS takes the given manually annotated training data, for a given relation in the form of (entity, relation, object) and its true score. The Feature Extractor makes the feature vector for each sample. The feature vectors for the all the samples are given for model training, which learns a relation-specific Relevance Model (RM) for the given two types of relations as given in the task: *profession* and *nationality*. During the prediction, for a given test triple, feature vector is produced by the Feature Extractor, the RM is then applied to it, that predicts a relevance score, which is an integer from the range from 0 to 7, which is also the output of the task.

### 3.1 Relevance Model (RM)

In our approach, we use the Logistic Ordinal Regression to learn the model from the given training data. The Logistic Ordinal Regression model, also known as the *proportional odds model*, was introduced by McCullagh *et al.* [12, 13], which is a generalized linear model which is designed to predict ordinal variables, that is, variables that are discrete (as in classification) but which can be ordered (as in regression). For all the training samples (entity, relation, object), we find the embeddings vector for entities and objects for every sample. As described in Fatma *et al.* [11], instead of simple words embeddings, we use word2vec entity embeddings of 100 dimensions generated using the open source tool[3], trained on the latest Wikipedia corpus[4]. The mentioned tool is a slightly modified version of word2vec where instead of treating the tokens of the entity/object as separate words, they are treated as a single unit while generating the word2vec embeddings. For example, the entire phrase *"United States of America"* is treated as a single entity and will have a single embedding. We use these embeddings to make well-designed features as explained in Section 3.2. We use the open source code[5,6], for training and learning the model. Just as the Logistic Regression models posterior probability $P(y = j|X_i)$ as the logistic function, the Logistic Ordinal Regression models the cumulative probability *P* as the logistic function, as defined below:

$$P(y \leq j|X_i) = \phi(\theta_j - w^T X_i) = \frac{1}{1 + exp(w^T X_i - \theta_j)}$$

where, $X \in \mathbb{R}^{nXp}$ denotes the input feature vectors, $w$, $\theta$ are the vectors to be estimated from the data, and $\phi$ is the logistic function as defined below:

$$\theta(t) = \frac{1}{1 + exp(-t)}$$

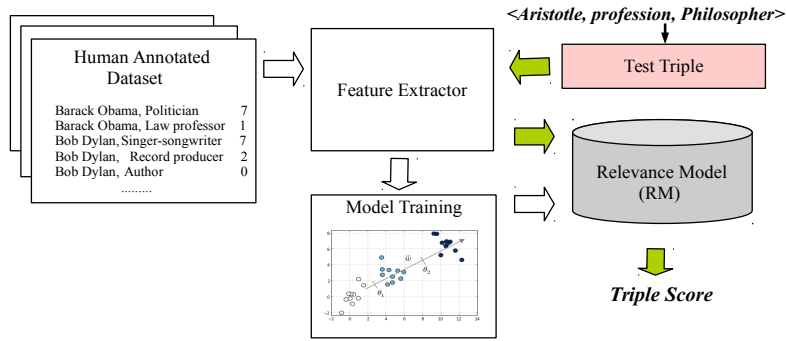

Figure 1: Celosia Triple Scorer (CTS) System Architecture

### 3.2 Feature Extractor

For a given triple (entity, relation, object), we compute four features described below, which helps in capturing the degree of relevance. We now describe the different features as follows:

- **Object-Entity Similarity:**

  For a given triple (entity, relation, object), we calculate the Cosine Similarity (C) between the person entity *e* and the object *o* embeddings vector as *C(e, o)*. The Cosine Similarity (C) is calculated as below :

  $$C(A, B) = cos\theta = \frac{A.B}{|A||B|} = \frac{\sum_{i=1}^{n} A_i B_i}{\sqrt{\sum_{i=1}^{n} A_i^2} \sqrt{\sum_{i=1}^{n} B_i^2}}$$

  where, $A_i$ and $B_i$ are components of vectors *A* and *B* respectively.

  It can be a useful feature in capturing the relatedness between the entity and object. For example, consider the triples *(Tupak Shakur, profession, Rapper)* and *(Tupak Shakur, profession, Social Activist)*. The cosine similarity between *Tupak Shakur* and *Rapper*, *C(Tupak Shakur, Rapper)*, is 0.77, which is much higher than the cosine similarity between *Tupak Shakur* and *Social Activist*, *C (Tupak Shakur, Social Activist)*, which is 0.36. Higher cosine similarity can be indicative of higher relevance.

- **Average Object-Page Entities Similarity (OPS):**

  For a given triple (entity, relation, object), we calculate the average cosine similarity between the object *o* and all the entities found in the Wikipedia page of the person entity *e*. For example, consider the triple *(Albert Einstein, profession, Physicist)*. In the Wikipedia page of the person entity *Albert Einstein*[7], we find that there exists entities like *"theoretical physicist"*,*"Nobel Prize in Physics"*,*"electromagnetic field"*, *"quantum mechanics"*, *etc.* which look semantically very related to the profession *Physicist*. It can be a very useful feature, because, if the given person entity *e* and the object *o* pair are very related, then the object *o* would usually also have higher similarity with the entities present in the Wikipedia page of the person entity *e* on an average.

| OPS Rank | Profession | OPS Score | True Score |
|---|---|---|---|
| 1 | Physicist | 0.67 | NA |
| 2 | Theoretical Physicist | 0.61 | 7 |
| 3 | Mathematician | 0.58 | 4 |
| 5 | Scientist | 0.57 | NA |
| 8 | Philosopher | 0.56 | 4 |
| 200 | Publisher | 0.07 | NA |

Table 1: OPS Rank of the professions for the person entity *Albert Einstein*

| OPS Rank | Nationality | OPS Score | True Score |
|---|---|---|---|
| 1 | Unites States of America | 0.39 | 7 |
| 3 | Germany | 0.35 | 3 |
| 4 | Austria | 0.34 | 5 |
| 100 | Ivory Coast | 0.19 | NA |

Table 2: OPS Rank of the nationalities for the person entity *Frederick Loewe*

---

[3]https://github.com/idio/wiki2vec
[4]https://dumps.wikimedia.org/enwiki/
[5]https://github.com/fabianp/mord
[6]http://fa.bianp.net/blog/2013/logistic-ordinal-regression/
[7]https://en.wikipedia.org/wiki/Albert_Einstein

We define the Average Object-Page Entities Similarity (OPS) for the entity *e* and the object *o* as follows:

$$OPS(e, o) = \frac{\sum_{i=1}^{N} C(o, e_i)}{N}$$

where, $e_i$ are entities in the Wikipedia page of person entity *e*, and *N* is the total number of entities in the Wikipedia page of the person entity *e*.

- **Object-Page Similarity Rank (OPS Rank):**

For the given triple (entity, relation, object), we make a list *S* of tuples ($o_i$, *OPS($o_i$,e)*) by calculating the *OPS* score, as described in 3.2, between every object $o_i$ belonging to set *O* and the given entity *e*, where the set *O* consists of all the professions or all the nationalities used in the task respectively. We then sort the list *S* according to the calculated *OPS* score, and find the rank of the object *o* of the given triple in the sorted list *S*. If the object *o* has a higher rank, it could be indicative of a higher relevance in comparison with other professions/nationalities. We define the OPS Rank below:

$$OPS\_Rank(e, o) = \text{Position of } o \text{ in the sorted list } S$$

where,

$$S(e) = \cup[(o_i, OPS(e, o_i))], \forall o_i \in O$$

Tables 1 and 2 show the OPS ranks for different professions and nationalities sorted by their OPS scores respectively. We observe that for the person entity *Albert Einstein*, the highly relevant professions, according to the human judgments or the True Score, like- *Theoretical Physicist, Scientist, Mathematician, etc.* also have higher OPS Ranks respectively. Similarly, for the person entity *Frederick Loewe*, the more relevant nationalities like *United States of America* and *Austria* have higher OPS Ranks respectively.

- **Object Mention:**

This feature is a slightly modified version of the *First* score as described in Bast *et al.* [8]. For a given triple (entity, relation, object), if the object *o* is mentioned in the Wikipedia page of the person entity *e*, irrespective of the position of mention, we give it a feature score 1. Many profession or nationality objects, even though mentioned at second, third or later positions might be very relevant with respect to the person entity *e*. Hence we give a score of 1 to every object mention, and use it as one of the features for our model. Whereas if there is no mention of the object *o*, we give a feature score of 0. This feature is particularly useful is scoring down the profession or nationality which is not relevant to the person entity *e* and would therefore not be mentioned in the Wikipedia page of the person entity *e*. For example, consider the Wikipedia page of the person entity *Aristotle*[8]. Many professions like *Basketball player, Carpenter, or Choreographer* which have no literal mention are most likely to be irrelevant.

---
[8]https://en.wikipedia.org/wiki/Aristotle

| Feature | Profession | Nationality |
|---|---|---|
| OPS Rank | 1.49 | 2.9 |
| Object Mention | 1.41 | 0.74 |
| Object-Entity Similarity | 1.21 | 0.73 |
| OPS | 0.14 | 0.1 |

Table 3: Feature weights learnt by the relevance model RM

### 3.3 Feature Importance

Table 3 shows the absolute weights of each feature learnt by the relevance model RM during the training phase in decreasing order of their value. These coefficients signify the importance of each feature. From the table, we see that OPS Rank is the most important feature of our model.

## 4. EXPERIMENTAL SETUP:

### 4.1 Dataset

We use the dataset provided by the challenge organizers [4]. It consists of a training data of: a) 515 triples for *profession* and b) 162 triples for *nationality*, respectively.

### 4.2 Baseline Systems

We implement two baseline systems in order to compare the performance of CTS:

- **First:** As described by Bast *et.al* [8], we find the first literal mention of a profession/nationality from the Wikipedia abstract of the person entity. That profession/nationality gets a score of 7 and all other profession/nationality gets a score of 0 respectively.

- **Logistic Regression Classification:** We implement this baseline in order to compare Logistic Regression Classification *vs* our approach CTS which implements Ordinal Logistic Regression Classification. This would help us to demonstrate the importance of order information which is being utilized by the latter to give better score predictions. In Section 3.1, we have described the mathematical description for both the approaches.

### 4.3 Evaluation Metrics

We use the standard evaluation metrics for the task such as *Accuracy*, *Average Score Difference* and *Kendall's Tau* as described below:

*Accuracy*: We calculate the percentage of triples for which the score computed by CTS differs from the score from the ground truth by at most 2 (i.e. $\delta = 2$).

*Average Score Difference (ASD)*: We calculate the average of the absolute difference of the relevance score of triples computed by CTS and their ground truth score respectively.

*Kendall's Tau:* For every unique entity from the given test-set triples, we compute the ranking of all triples with respect to the subject and relation according to the scores computed by: a) CTS and b) the score from the ground truth. We then compute the difference of the two rankings using Kendall's Tau [14] which measures the correlation between two ranked lists.

## 5. RESULTS AND DISCUSSION

Table 4 shows the 5-fold cross validation evaluation comparison of the baselines and the CTS trained on the best model parameters.

| Method | Profession | | | Nationality | | |
|---|---|---|---|---|---|---|
| | Accuracy ($\delta = 2$) | ASD | Kendall's Tau | Accuracy ($\delta = 2$) | ASD | Kendall's Tau |
| *First* | 0.43 | 3.50 | 0.69 | 0.32 | 4.14 | 0.58 |
| *Logistic Regression* | 0.64 | 2.32 | 0.43 | 0.71 | 1.81 | 0.64 |
| *CTS* | **0.69\*** | **1.74\*** | **0.35\*** | **0.77\*** | **1.55\*** | **0.42\*** |

Table 4: 5-Fold Cross Validation Results: CTS performance in comparison with baseline approaches. Results marked with a * were found to be statistically significant with respect to the nearest baseline at 95% confidence level ($\alpha = 0.05$) when tested using a two-tailed paired t-test.

| Relation | Accuracy ($\delta = 2$) | ASD | Kendall's Tau |
|---|---|---|---|
| *Profession* | 0.71 | 1.80 | 0.33 |
| *Nationality* | 0.75 | 1.71 | 0.42 |
| *Average Score* | 0.73 | 1.75 | 0.36 |

Table 5: Overall Test Results of the CTS

The results show that the CTS performs significantly better than the baseline models. There was an improvement of 7.81% in accuracy for professions and 8.45% in accuracy for nationality from the nearest baseline of the Logistic Regression based model. Also, there was an improvement in the ASD score of 25% for professions and 14.36% for nationality. The Kendall's tau score showed an improvement of 18.6% for profession and 34.37% for nationality. This signifies how the order information which we use in CTS has a significant impact in comparison to simple Logistic Regression based approach.

Table 5 shows the overall experimental results of CTS on the test set which was used in the evaluation by the challenge organizers [4]. We achieve an overall accuracy score of 0.73, an ASD score of 1.75, and a Kendall's tau score of 0.36.

## 6. CONCLUSION AND FUTURE WORK

We used a supervised machine learning approach of Ordinal Logistic Regression based classification model which *learns* the notion of relevance from the human annotated training data and uses it to predict relevance scores for other triples of similar type relations. We achieve an overall accuracy of 0.73 in the challenge task. We carefully designed new features for the task. From our experiments, we observe that our approach significantly out performs the baselines. This signifies that even with the few but well-sought features which we designed like the OPS rank, Object Mention, etc., the model turned out to be quite successful in solving the task, although we had a very limited number of training data samples. In future, we are planning to do more feature engineering, and apply new approaches. We also plan to try alternate feature representation for entities and objects.